\documentclass[journal,onecolumn,]{gji}
\usepackage{timet,color}
\usepackage[urlcolor=blue,citecolor=black,linkcolor=black]{hyperref}

\usepackage{xcolor, soul}
\sethlcolor{yellow}

\usepackage{amssymb}
\usepackage{amsmath}
\usepackage{bm}

\usepackage{graphicx}

\usepackage{epstopdf}
\AppendGraphicsExtensions{.tiff}
\usepackage{float}

\usepackage[T1]{fontenc}
\usepackage[english]{babel}

\usepackage[authoryear]{natbib}

\bibpunct[, ]{(}{)}{,}{a}{}{,}%
\usepackage{setspace} 

\title{3D Muographic Inversion in the Exploration of Cavities and Low-density Fractured Zones}

\author[L. Balázs et al.]{
\parbox{\linewidth}{
László Balázs,$^{1,2}$ 
Gábor Nyitrai,$^{1,3}$\thanks{E-mail: nyitrai.gabor@wigner.hu} 
Gergely Surányi,$^{1}$ 
Gergő Hamar,$^{1}$ \\
Gergely Gábor Barnaföldi,$^{1}$ 
and Dezső Varga$^{1}$ }%
\\
$^{1}$Wigner Research Center for Physics, Konkoly-Th. M. u. 29-33, 1121 Budapest, Hungary\\
$^{2}$Department of Geophysics and Space Science, Eötvös Loránd University, Pázmány P. s. 1/C, 1117 Budapest, Hungary\\
$^{3}$Budapest University of Technology and Economics, Műegyetem rkp. 9, 1111 Budapest, Hungary
}

\date{Received 2023 February 16; in original form 2023 February 16}
\pagerange{\pageref{firstpage}--\pageref{lastpage}}

\begin{document}

\label{firstpage}

\maketitle

\begin{summary}
Muography is an imaging tool based on the attenuation of cosmic muons to observe the density distribution of large objects, such as underground caves or fractured zones. Tomography based on muography measurements -- that is, three dimensional reconstruction of density distribution from two dimensional muon flux maps --  brings up special challenges. The detector field of view covering must be as balanced as possible, considering the muon flux drop at higher zenith angles and the detector placement possibilities. The inversion from directional muon fluxes to 3D density map is usually underdetermined (more voxels than measurements) which can be unstable due to partial coverage. This can be solved by geologically relevant Bayesian constraints. The Bayesian principle results in parameter bias and artifacts. In this work, the linearized (density-length based) inversion is applied, the methodology is explained, formulating the constraints associated with inversion to ensure the stability of parameter fitting. After testing the procedure on synthetic examples, an actual high quality muography measurement data set from 7 positions is used as input for the inversion. The result demonstrates the tomographic imaging of a complex karstic crack zone and provides details on the complicated internal structures. The existence of low density zones in the imaged space was verified by samples from core drills, which consist altered dolomite powder within the intact high density dolomite. 
\end{summary}


\section{Introduction} 
\label{sec:Introduction}

Cosmic radiation on the surface of the Earth is a natural phenomenon known for a century, based on the pioneering discovery by~\citet{Wulf_1909}. Shortly after the cosmic origin was demonstrated by the balloon measurements of Victor Hess -- by the observed increasing radiation in the higher atmosphere~\citep{Hess_1912}. Since then, several key properties of cosmic rays have been discovered, showing that most of those reaching Earth surface are muon particles: collision cascade and subsequent decay products of primary cosmic rays of Galactic origin. Nature is rather kind to us on Earth: unlike other planets in the Solar System~\citep{Kedar_2013,LeoneMars}, the atmosphere is sufficiently thick to filter most of the hadrons, and thin enough to let through most of the produced muons. This allows one to precisely quantify the muon flux 
at any point on the surface of the Earth or up to kilometer deep underground, with negligible time variation for the purposes of this paper. 

Imaging with cosmic muons, or "muography" in short, was pioneering in the 60s, first applied for archaeology by~\citet{Alvarez}. Ever since, this emerging scientific field has a long history~\citep{Kaiser_2019}, an overview of which is well covered in a recent monograph by~\citet{AGU_book}. One of the basic setting is "underground muography": the flux reduces strongly and in a well predictable manner as a function of material crossed, similar to X-ray radiography. The flux is nearly precisely dependent only on the integrated density along the measurement line, and the zenith angle~\citep{Lesparre_2010}. In fact, the muon travels along a nearly straight line up to the point of stopping~\citep{Olah_2019_limits},
 which allows a directional measurement with a sufficiently precise muon tracking detector. It is also worth mentioning here, that not only the attenuation of muons can be utilized for imaging. For example, the muon scattering indicates density and atomic number distribution inside an examined volume by measuring the direction of the particles before and after the target. This method was first proposed by a group from the Los Alamos National Laboratory \citep{scattering_2, scattering_3}, and recently an overview of similar studies was given by \citet{scattering_1}.

The aim of underground muography measurement is to investigate the "target", that is, the density distribution in the volume above the detector. In order to map the three dimensional structure, multiple measurement views of two dimensional measurements (tomography) may be invoked. This is inherently an inversion problem, since the muon flux is related to a complicated integral of the density. 

This work demonstrates practical tomographic inversion of actual muographic data, and shows the structural determination of a three-dimensional (3D) crack zone by multi-view muography. Radiographic approach to image inhomogeneous density zones with muography has been applied earlier by~\citet{Tanaka_2007, Tanaka_2011, Tanaka_2020, Miyamoto_2017, Olah_2021} among others. A tomographic measurement is rather complex~\citep{AGU_Miyamoto_chap},
 and one needs not only high statistics data, but systematic errors need to be well controlled from the multiple views. The inversion needs to use some existing information, that is, a Bayesian approach, and may invoke other measurement techniques such as gravity~\citep{AGU_Nishiyama_chap, Barnoud_2019, Guardincerri_2017, Cosburn_2022} or electric resistivity~\citep{Lesparre_2012}. Muon tomography has been successfully applied earlier, such as to identify underground density increase by an ore body~\citep{canada_2018} and decrease by cavities~\citep{florence_2022, naples_2019, Liu_2023}, or to image the internal structure of a volcanic cone~\citep{Nagahara_2022} or nuclear reactor~\citep{Procureur_2023}. The present paper describes the adaptation of a maximum likelihood inversion method with the combination of geologically relevant Bayesian constrains for multi-view muographic measurements. The method includes uncertainty propagation, quantifications for focus zone determination, and synthetic data test. 
 
\begin{figure*}
    \centering
    \includegraphics[height=6cm]{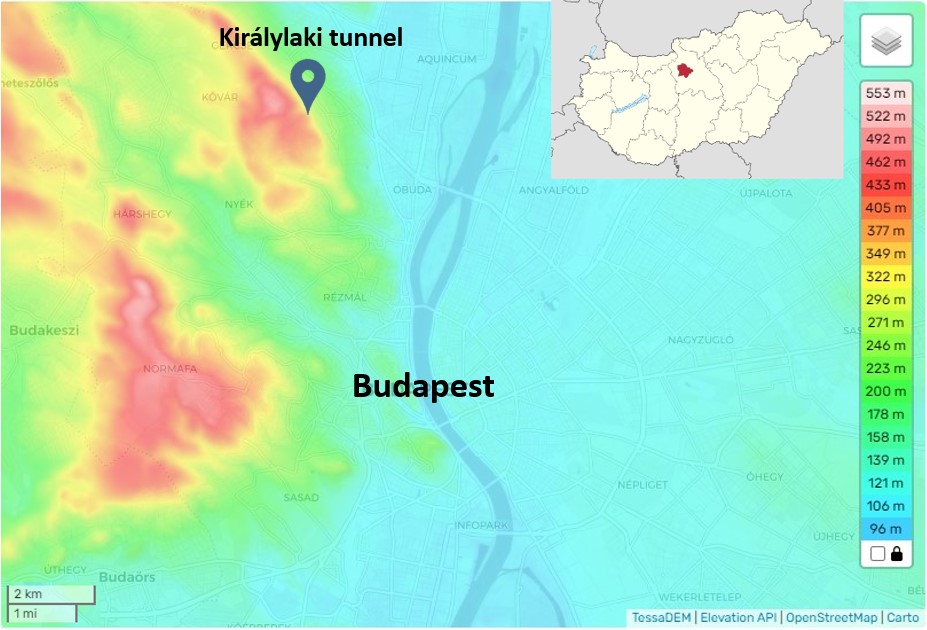} 
    \caption{
Geograpic location of the measurements in Budapest. The horizontal tunnel is oriented towards the west, with increasing overburden along its length.
}
    \label{fig:topo}
\end{figure*}

 The applicability for 3D density estimation is demonstrated for the first time on a high resolution muography survey of a karstic underground crack zone at shallow depth (40--60~m), and validated by drill samples. The measurements were performed in the Királylaki  tunnel systems near Budapest, as indicated in Fig.~\ref{fig:topo}, along a horizontal tunnel. The location was chosen due to the possible convenient geometry of the measurements along a straight line, but at the same time expecting a complicated density structure above the tunnel.
 

\section{Linearized muon-tomographic density distribution reconstruction} 
\label{sec:Linearized}

In linearized tomographic inversion the initial data system is the line integral of the density (density-lengths) obtained by transforming the directionally recorded pulse count rates of the muon detectors. More precisely, one quantifies the "integrated energy loss" along the given direction, however, this has little (or if needed quantifiable) dependence on material composition~\citep{Lesparre_2012}. In this approximation, the response of the volume of interest (direct problem) can be described as a linear transformation of the discretized density distribution of the surveyed geological object, where the transformation is a function of the measurement geometry alone. To estimate the correct density distribution, the error distribution of count rates (related to the flux) must also be transformed.

\subsection{Forward problem: the geometric model} 

\begin{figure*}
    \centering
    \includegraphics[height=7cm]{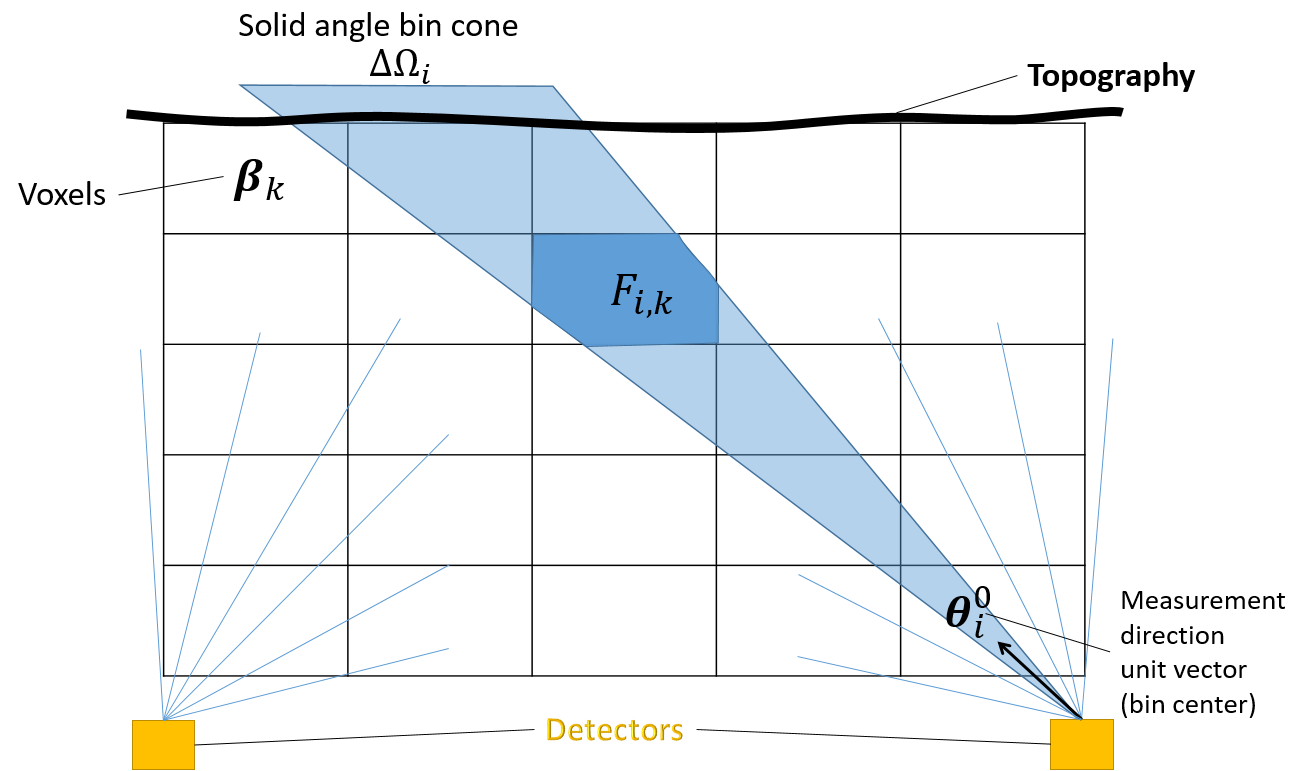} 
    \caption{
Schematic representation of the relevant variables and their geometrical relations. The measurement is done in a finite solid angle bin, which intersects a specific voxel. The mean path length in the voxel is encoded in the $\textbf{F}$ matrix. (The voxel size and the solid angle cone is exaggerated relative to a typical configuration for the sake of clarity).
}
    \label{fig:schematic}
\end{figure*}

In the case of absorption muon tomography, one tries to estimate the density distribution $(\rho(\mathbf{r}),\mathbf{r} \in V)$ of a simply connected continuous domain $(V)$ (partially bounded by the known surface topography). For that, multi-view radiographic muon transmission data are collected from the domain, defined as density-length 
\begin{equation}
    \gamma = \int_L \rho(\lambda) \, d\lambda, 
\end{equation}
where $\lambda$ is the length measure along a muon trajectory $L$, and the aim is to determined the density distribution by solving the inverse problem, similarly to X-ray based computed tomography (CT). The muon transmission derives from the attenuation of cosmic muons through the medium, depending basically on the density properties of the medium due to energy loss. This means, that with increasing density-length, the rate of cosmic muons decreases and the muon spectrum hardens. The initial -- energy ($E$) dependent -- muon flux spectrum on the surface of the Earth ($\varphi_0(\boldsymbol{\theta},E)$) depends also on the zenith angle of the arriving muon (its direction norm vector will be denoted by $\boldsymbol{\theta}$). In this paper, the method by~\citet{Guan_2015} has been used to approximate $\varphi_0$.
The muon flux measurements are performed below the domain $V$ of interest at specific locations, that is, sampled on a few points of the domain edge (in the detector locations, indicated as rectangles in the bottom of Fig.~\ref{fig:schematic}.) by the direction sensitive measurements. In these points, a muon tracking detector registers count rates generated by muons with a surface energy greater than a threshold value ($E_{\mathrm{min}}$) which required to survive through the medium. The actual measured data is a vector of the muon count number $\textbf{y}$, which can be expressed as count rate $\dot{\textbf{y}} =  \textbf{y} / \Delta t$, where $\Delta t$ is the measurement live time. In turn, the count rate can be normalized to flux by dividing with $\eta$:~the product of detector efficiency, acceptance (effective sensitive area).

Muons measured at different positions and discretized (binned) by direction angle represented by direction unit vectors $\boldsymbol{\theta}^{0}_i$ (bin center of $i$\textsuperscript{th} measurement) with disjoint solid angle ranges ($\Delta \Omega_i$) of bins, on which the detector integrates the muon arrivals. The theoretically expected muon count rates in the measurements are:


\begin{equation}\label{eq:flux}
    \Psi_i =  \int\displaylimits_{\boldsymbol{\theta} \in \Delta \Omega_i} \int\displaylimits_{E \geq E_{\mathrm{min}}(\gamma_i)} \eta_i \varphi_0(\boldsymbol{\theta}, E)dE d{\Omega} \approx 
    \eta_i \Delta \Omega_{i}
    \int\displaylimits_{E \geq E_{\mathrm{min}}(\gamma_i)}\varphi_0(\boldsymbol{\theta}^{0}_i, E)dE 
\end{equation}
here $\boldsymbol{\theta}$ is the measured muon direction, $\textbf{r}_i$ is the detector position of $i$\textsuperscript{th} measurement, the index $i \in [1..N]$ where $N$ is the total number of measurement bins, and $\gamma_i$ the density-length derived from the modeled medium. The expected muon count rate is useful for significance and detection limit calculations. With sufficiently small angular bins, one can use the efficiency calculated in the bin center, and the integration over solid angle translates to a multiplication with the bin size $\Delta \Omega_i$.


The association between the density-length and the muon count rate can be derived from equation \eqref{eq:flux} through the $E_{\mathrm{min}}(\gamma)$. The minimum required energy for a muon to survive after energy deposition in a given density-length is well known in the literature, we use the method by~\citet{Lesparre_2010} for this calculation.


However the direct measurement data is muon count number ($\dot{\textbf{y}}$), it is preferable to make two steps further to produce the input data for inversion.
The first one is to determine the measured (energy integrated) flux $\Phi^m$, 
\begin{equation}\label{eq:fluxmeasured}
    \Phi_i^m (\boldsymbol{\theta}^{0}_i,\textbf{r}_i)  = \frac{\dot{\textbf{y}}}{\eta_i \Delta \Omega_{i}}.
\end{equation}
From here on, the superscript \textit{m} denotes quantities derived from measurements. 
Even though muon flux is derived directly from measured data, it is not easy to interpret, therefore it is rather transformed to density-length in a second step. This transformation requires $\varphi_0$ as input, as well as the above quoted $E_{\mathrm{min}}(\gamma)$ dependence, resulting in the (implicitly expressed) measured density-length $\gamma^m$:

\begin{equation}\label{eq:fluxdensitylength}
    \Phi_i^m =  
    \int\displaylimits_{E \geq E_{\mathrm{min}}(\gamma_i^m)}\varphi_0(\boldsymbol{\theta}^{0}_i, E)dE.
\end{equation}

Using $\boldsymbol{\gamma}^m$ instead of $\dot{\textbf{y}}$ or $\Phi^m$ (count rate or measured flux) do not only result in an easier interpretation by geoscientists, but linearizes the inversion problem since density-length is an additive quantity.


%

The directions of the measured muon trajectories are within the angular bin represented by its central direction vector $\boldsymbol{\theta}_i^0$. The angular bins are  not always negligibly small, which means that one must average over all the directions within the $\Delta \Omega$ bin size. The density-length can be readily calculated as the integral of the density distribution $\rho(\boldsymbol{\textbf{r}})$ along (the averaged) trajectory.

%

For the tomographic reconstruction, the continuous density distribution of the volume $V$ is expressed in a finite dimensional basis (in the simplest case a grid, covering $V$) as a form of parameter discretization. The grid elements $\boldsymbol{\beta}_k$  are non-overlapping volume elements which provide a disjoint coverage of the domain $V$ under study ($k \in [1..K]$, where $K$ denotes the number of volume elements). These grid elements (expressed as a vector) can be used to generate orthogonal basis functions to describe the density distribution:
\begin{equation} \label{eq:betadef}
    \beta_k(\textbf{r}) = \Biggl[_{\, 0; \, \textbf{r} \notin \boldsymbol{\beta}_k }^{\, 1; \, \textbf{r} \in \boldsymbol{\beta}_k} \ ,
\end{equation}

that is, the $\beta_k(\textbf{r})$ functions take the value of 1 only within the $k$-th volume element, and zero outside the element. This formalism is indicated schematically in Fig.~\ref{fig:schematic}. An element of the density vector ($\boldsymbol{\rho}$) is therefore constructed by the discretization of density distribution $\rho_k = \rho(\textbf{r})\cdot \beta_k(\textbf{r})$. 
%
%
Let $F_{i,k}$ denote the path length of (averaged) muon trajectories within grid elements, approximately indicated by the shaded cross section area in Fig.~\ref{fig:schematic}. This $\textbf{F}$ is the Jacobian matrix, transforming between the discretized density and density-length (index $i$ runs through the measurement lines, while index $k$ refers to the grid elements). The matrix includes all the geometric information needed for the inversion. Adding up the path lengths, weighted with the density, results the vector of density-lengths: 
\begin{equation}
    \boldsymbol{\gamma} = \textbf{F} \boldsymbol{\rho}.
\end{equation}
Given a suitable (error model dependent) metric, the distance from the measurement (vector norm of $\Delta \boldsymbol{\gamma} = \boldsymbol{\gamma}^m - \textbf{F} \boldsymbol{\rho} $, where $\gamma_i^m$ vector elements are derived from Eq.~\ref{eq:fluxdensitylength}) will be the basis for the reconstruction. According to Fig.~\ref{fig:schematic} the topography defines the upper boundary of the volume of interest, that is, the path lengths and density distribution are calculated only below the topography (zero above).

\subsection{Inverse problem and related errors} 
\subsubsection{The complex error model associated with tomographic inversion} 

Muon tomography is characterized by imaging over a relatively restricted angular range (looking only upwards), which is therefore a cone beam type limited angle transmission tomography. The one-sided arrangement implies the possibility of distortion of the fitted distribution along the projection lines and associated artifacts. The number of measurement points, angular resolution and the coverage (projection line network density) determine the possible resolution for the reconstruction, i.e. the finest grid subdivision that can be applied  \citep{Tanaka_2017}. 

The method of parameter fitting (density reconstruction) is determined by the error model associated with the imaging. It also determines the functional (quadratic form, $Q$) to be minimized during parameter fitting process. The error of the measured data can be considered as Poisson distribution, of which the Maximum Likelihood (ML) principle leads (approximately) to the weighted least squares (WLS) fitting, to minimize the following functional with the measured data ($\textbf{y}$) and theoretical muon count rates ($\boldsymbol{\Psi}$):
\begin{equation}
    Q_{\mathbf{y}}(\boldsymbol{\gamma}) = \sum\displaylimits_{i=1}^{N} W_{\mathbf{y},i,i}(\mathrm{y}_i - \Psi_i(\gamma_i)\Delta t)^2; \quad W_{\mathbf{y},i,i} \approx \frac{1}{\mathrm{y}_i} \ , 
\end{equation}
where $\textbf{W}_{\mathbf{y}}$ the weight matrix (inverse of measurements covariance matrix), for the $N$ number of measurements.

As explained above, ($\textbf{y} \rightarrow \boldsymbol{\gamma}^m$) transformation simplifies the tomographic inverse problem, which also transforms the original error distribution. Changing of data variances requires the modification of weight ($\textbf{W}_\gamma$), taking care of error propagation. Since the transformation is non-linear, it causes a small bias ($\delta\gamma_i = [\mathbb{E},\Psi^{-1}]y_i$) which is negligible at count rates~\citep{Szatmary_2002} relevant for muography (counts per bin above 100). The functional associated to the fit after the transformation becomes linear:
\begin{equation}
    Q_\rho = (\boldsymbol{\gamma}^m - \textbf{F}\boldsymbol{\rho})^T\textbf{W}_\gamma(\boldsymbol{\gamma}^m - \textbf{F}\boldsymbol{\rho}) \ .
\end{equation}

The weight matrix elements are related to the transformation between flux and density-length:
\begin{equation}
    W_{\gamma,i,i} = (\eta_i  \Delta t_i)^2 \left( \frac{\partial\Phi_i^m}{\partial\gamma_i} \right)^2  \frac{1}{\mathrm{y}_i } \ .
\end{equation}
The density-length covariance matrix ($\textbf{C}_\gamma$) is assumed to be diagonal, neglecting the small-scale data correlation due to the transformation. Expressing with the weight matrix:
\begin{equation}
    \textbf{C}_\gamma = \sigma^2_\gamma \textbf{W}_\gamma^{-1} \ , 
\end{equation}
where $\sigma_\gamma^2$ is the calibration factor for the posterior $\textbf{C}_\gamma$ which can be estimated from the post-fitting value of normalized $Q_\gamma$. 

Up to now, the grid structure is arbitrary, defined by Eq. \ref{eq:betadef}. This grid may be adapted to the geometry of the domain to be mapped, using the direction-dependent weight of the measurements:
\begin{equation}\label{eq:omega}
    w_k = \sum\displaylimits_{i=1}^{N}F_{i,k} W_{\gamma, i, i} \ .
\end{equation}

The weights can be used to filter out blind spots (low weight voxels) in a given measurement configuration.
Increasing the grid size can locally improve the weight of a grid element, but at the same time reduces the variance of the density estimate of the grid element, degrades the spatial resolution and, with varying density, the fitted density value may become less and less representative of the environment. In the case of cavity exploration, it is generally assumed that cavities are located at zero density in a rock mass with a relatively accurately known density. Increasing the size of grid elements may blur and reduce the density contrast.
It is worth noting that the topographic error appears as a parameter error, a small near-surface density anomaly after the fit. The data system can also be checked for the detectability of possible density anomalies before the fit. The feasibility conditions for muographic surveys has been examined in multiple papers, eg. in~\citet{Leone_rspa} or in ~\citet{Lesparre_2010}.

\subsubsection{Mathematical background of inversion} 

A typical muon tomography of cavity exploration sometimes requires a resolution of up to one meter, if possible, and a corresponding grid spacing. When designing the grid, it should be borne in mind that the resolution deteriorates as the distance from the measurement site increases (due to the angle uncertainty). The large number of grid elements (number of parameters) may lead to underdetermination of the parameters to be fitted (ambiguity, high parameter correlation) in parts of the mapped range. To make this task mathematically tractable, but also geologically relevant, Bayesian maximum {\it a posteriori} probability (MAP) principle was used to set up the fitting criterion. This means that the likelihood function $L(\boldsymbol{\gamma},\boldsymbol{\rho})$ is also multiplied by the {\it a priori} density function for the parameter distribution $p(\boldsymbol{\rho})$. Thus, the functional $Q_\gamma$ from the ML principle is complemented by another quadratic form $Q_\rho$ for the parameters to be fitted \citep{Menke_2018, Tarantola_1987}:
\begin{equation}
    Q^{(0)} = Q^{(0)}_\gamma + Q^{(0)}_\rho = (\boldsymbol{\gamma}^m - \textbf{F} \boldsymbol{\rho})^T \textbf{W}_\gamma (\boldsymbol{\gamma}^m - \textbf{F} \boldsymbol{\rho}) + (\boldsymbol{\rho} - \boldsymbol{\rho}^{(0)})^T \textbf{W}_\rho^{(0)} (\boldsymbol{\rho} - \boldsymbol{\rho}^{(0)}) \ ,
\end{equation}
where $\textbf{W}_\rho^{(0)} = (\textbf{C}_\rho^{(0)})^{-1}$ matrix is the inverse of the covariance matrix of the {\it a priori} distribution in Bayes' theorem. The above functional provides a dual metric for fitting the density vector. The first part is the criterion for the measurement space, the second part is the {\it a priori} requirement for the parameter space: $\boldsymbol{\rho}^{(0)}$ centered on the assumed parameter distribution.
Setting $\boldsymbol{\rho}^{(0)}$ to a constant value, it corresponds to assuming a prior with Gaussian distribution around the well-defined solid rock density. 
Diagonal elements of $\textbf{W}_\rho^{(0)}$ can be seen as effective error terms for the a priori distribution~\citep{Tarantola_1987} whereas the matrix can include various forms of damping or smoothing. Since the statistical errors are more complicated to evaluate in that (non-diagonal matrix) case, and the results seemed stable in the setting of muographic inversion, $\textbf{W}_\rho^{(0)}$ was set to be proportional to a unity matrix. The normal equations associated with parameter fitting:
\begin{equation}
    \partial_\rho Q^{(0)} = -\textbf{F}^T \textbf{W}_\gamma \boldsymbol{\gamma}^m + \textbf{R} \boldsymbol{\rho} + \textbf{W}_\rho^{(0)} ( \boldsymbol{\rho} - \boldsymbol{\rho}^{(0)}) = \textbf{0} \ ,
\end{equation}
where $\textbf{R} = \textbf{F}^T \textbf{W}_\gamma \textbf{F}$ notation has been applied to the symmetric quadratic matrix in the formula. Hence the first order estimate of the density distribution \citep{Tarantola_1987}:
\begin{equation}\label{20}
    \boldsymbol{\rho}^{(1)} = (\textbf{R} + \textbf{W}_\rho^{(0)})^{-1} (\textbf{F}^T \textbf{W}_\gamma \boldsymbol{\gamma}^m + \textbf{W}_\rho^{(0)} \boldsymbol{\rho}^{(0)}) \ .
\end{equation}
The covariance matrix associated with the expected value (denoted by $\mathbb{E}$) of the density distribution vector variation ($\delta\boldsymbol{\rho}$) in equation (\ref{20}) can be derived as follows: 
\begin{multline}
    \textbf{C}_\rho^{(1)} = \mathbb{E} \left(\delta\boldsymbol{\rho}^{(1)} (\delta\boldsymbol{\rho}^{(1)})^T\right) = \\
     \mathbb{E} \left[ (\textbf{R}+\textbf{W}_\rho^{(0)})^{-1} (\textbf{F}^T \textbf{W}_\gamma \delta\boldsymbol{\gamma}) (\delta\boldsymbol{\gamma}^T \textbf{W}_\gamma \textbf{F}) (\textbf{R} + \textbf{W}_p^{(0)} )^{-1} \right] + \\
     \mathbb{E} \left[ (\textbf{R}+\textbf{W}_\rho^{(0)})^{-1} (\textbf{F}^T \textbf{W}_\gamma \delta\boldsymbol{\rho}) (\delta\boldsymbol{\rho}^T \textbf{W}_\gamma \textbf{F}) (\textbf{R} + \textbf{W}_p^{(0)} )^{-1} \right],
\end{multline}
where the estimated density vector variation is a function of measured density-length variation and the prior density variation. The above formula combines the effect of these two sources of error in a weighted way \citep{Menke_2018}.  After some rearrangement:
%
\begin{equation}
    \textbf{C}_\rho^{(1)} = \left(\textbf{R} + \textbf{W}_\rho^{(0)}\right)^{-1}  .
\end{equation}
%

The Bayesian assumption introduces a bias with respect to the first-order asymptotically unbiased ML (Maximum Likelihood) estimate. To estimate the bias, we can calculate the relationship between the true and the estimated densities:
\begin{equation}
    \mathbb{E} \left[ \boldsymbol{\rho}^{(1)} \right] = \left(\textbf{R} + \textbf{W}_\rho^{(0)}\right)^{-1} \left( \textbf{F}^T \textbf{W}_d \textbf{F} \boldsymbol{\rho}_{\mathrm{real}} + \textbf{W}_\rho^{(0)} \boldsymbol{\rho}^{(0)} \right) \ .
\end{equation}
Hence the expected value of bias ($\textbf{b}^{(1)} = \mathbb{E} \left[ \boldsymbol{\rho}^{(1)} - \boldsymbol{\rho}_{\mathrm{real}}\right]$):

\begin{eqnarray}
    \textbf{b}^{(1)} = &   \left( \textbf{R} + \textbf{W}_\rho^{(0)} \right)^{-1} \left( \textbf{F}^T \textbf{W}_d \textbf{F} \boldsymbol{\rho}_{\mathrm{real}} + \textbf{W}_\rho^{(0)} \boldsymbol{\rho}^{(0)} \right) - \boldsymbol{\rho}_{\mathrm{real}}
\\
     = & \left(\textbf{R} + \textbf{W}_\rho^{(0)}\right)^{-1} \left( (\textbf{F}^T \textbf{W}_d \textbf{F} - \textbf{R} - \textbf{W}_\rho^{(0)})\boldsymbol{\rho}_{\mathrm{real}} + \textbf{W}_\rho^{(0)} \boldsymbol{\rho}^{(0)}  \right) 
\\
     = & \left( \textbf{R} + \textbf{W}_\rho^{(0)} \right)^{-1} \textbf{W}_\rho^{(0)} \left( \boldsymbol{\rho}^{(0)} - \boldsymbol{\rho}_{\mathrm{real}} \right).
\end{eqnarray}

The bias occurs just at the cavities reducing the indication. Non-diagonal elements of $\textbf{R}$ contribute to the appearance of a smaller artifact. Given the histogram of the estimated parameters, a critical level can be defined for the separation of cavities during post-processing. Recall also that almost any regularized version of the method of least squares is equivalent to some form of Bayesian estimation.

\subsubsection{Reducing the inversion problem from 3D to 2+1D}\label{2+1D}

Inversion of a muon tomography measurement is an inherently three dimensional problem. In special cases, such as when the measurement locations are along a line -- just as it happened for those to be presented in this paper -- the configuration reduces to an independent sum of 2 dimensional problems, to planes containing the measurement line. 
The inversion planes never intersect other than the measurement line, therefore both the measured density-length values, as well as the resulting density distributions are uncorrelated and independent. Inversion is solved in each plane independently, and the 3 dimensional density distribution is derived by the projection of the solutions to the voxel base of the examined space.  The general formulation is fully valid, which means that one can expect the same artifacts in the inversion plane due to biases like in a "fully 3D" inversion.



\section{Simulated measurements - preliminary tests} 

\begin{figure*}
    \centering
    \resizebox{1\textwidth}{!}{
    \includegraphics[height=5cm]{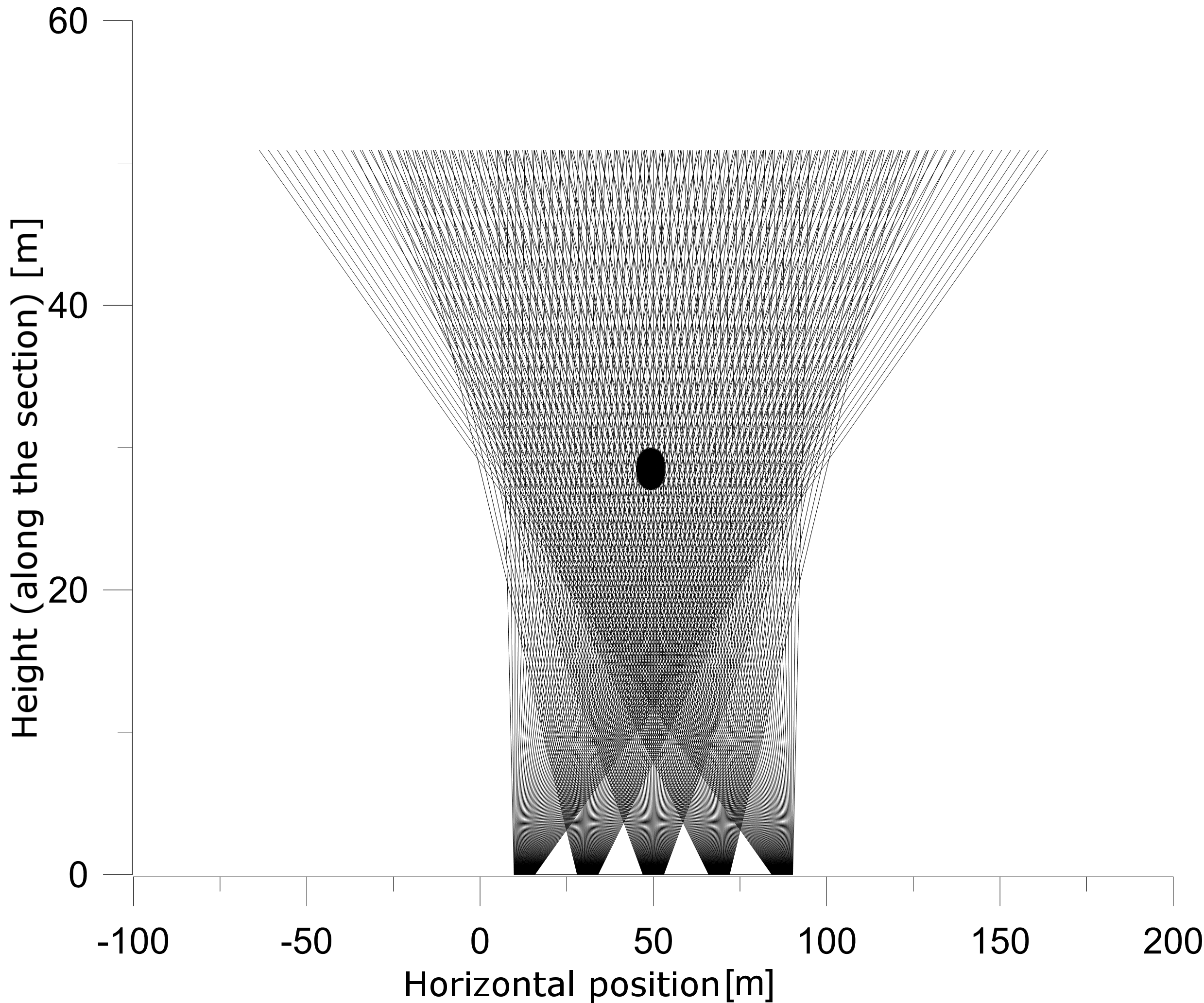}
    \includegraphics[height=5cm]{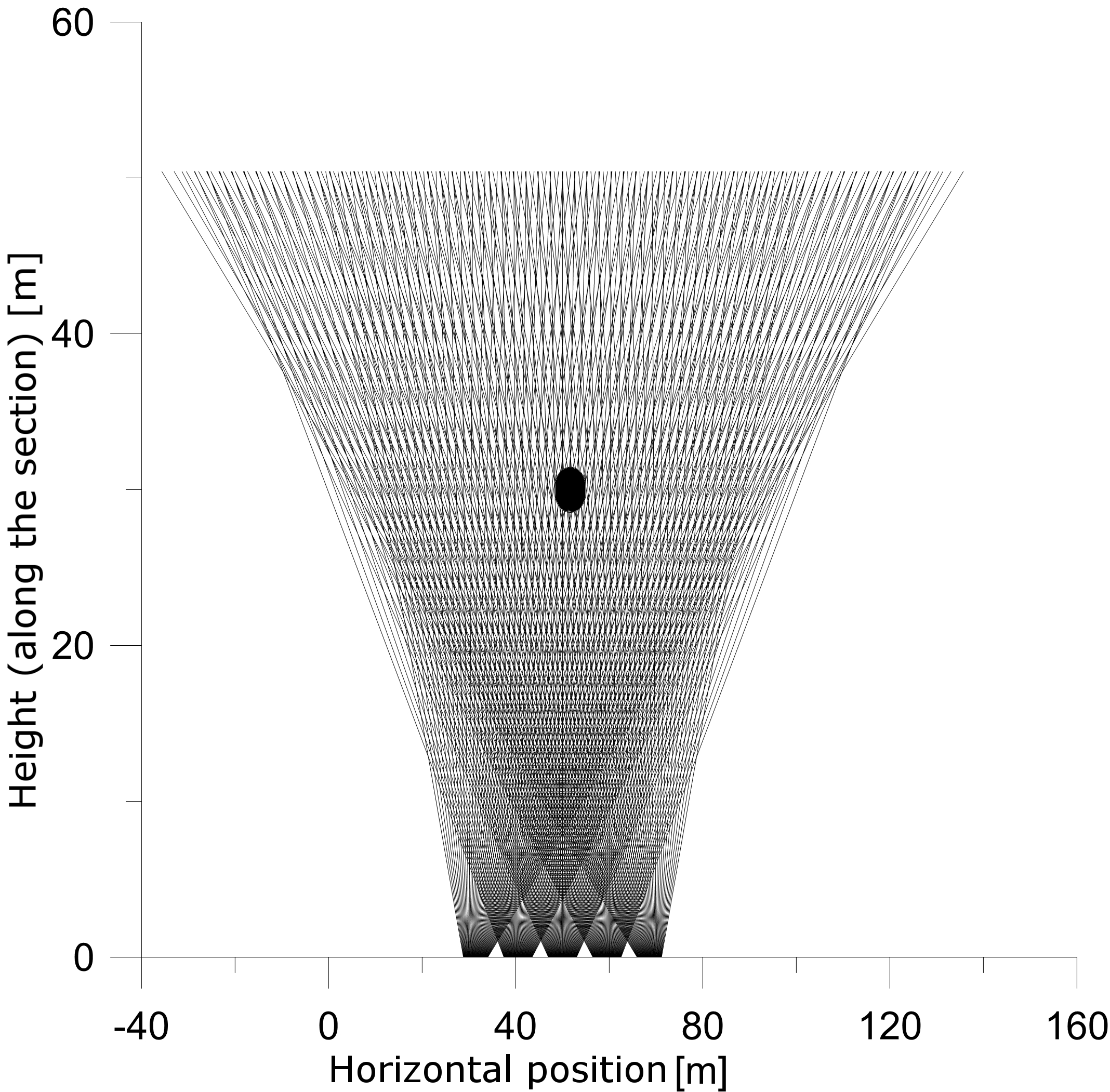} }
    \caption{
Projection lines of two measurement layouts under investigation: with different measurement points and detector tilt angles. The known cavity is located at the center. The voxel resolution is 1 m in horizontal and 0.5 m in vertical direction. The scale of axes are not the same.
}
    \label{fig:1}
\end{figure*}

\begin{figure*}
    \centering
    \resizebox{1\textwidth}{!}{
    \includegraphics[height=5cm]{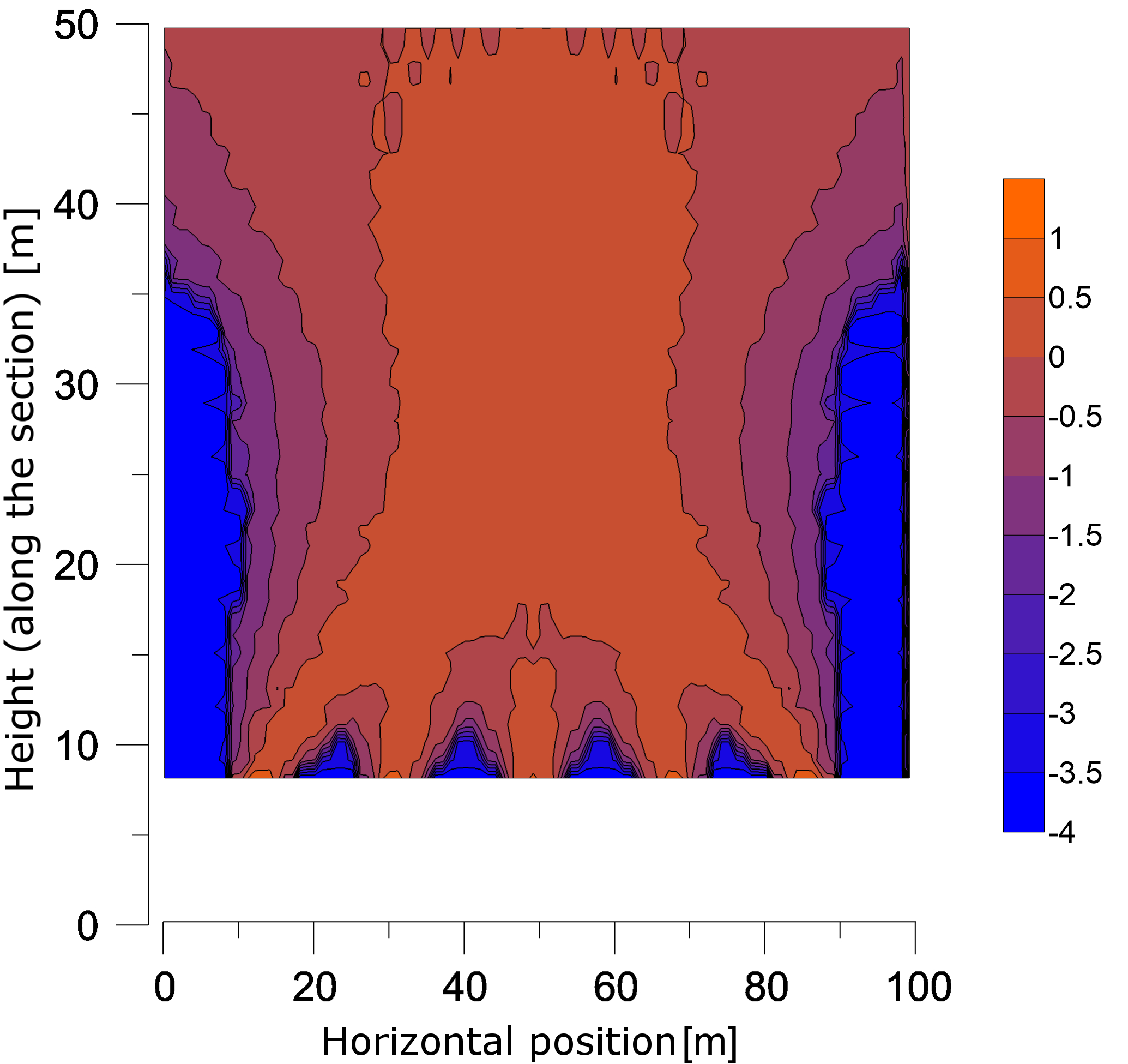}
    \includegraphics[height=5cm]{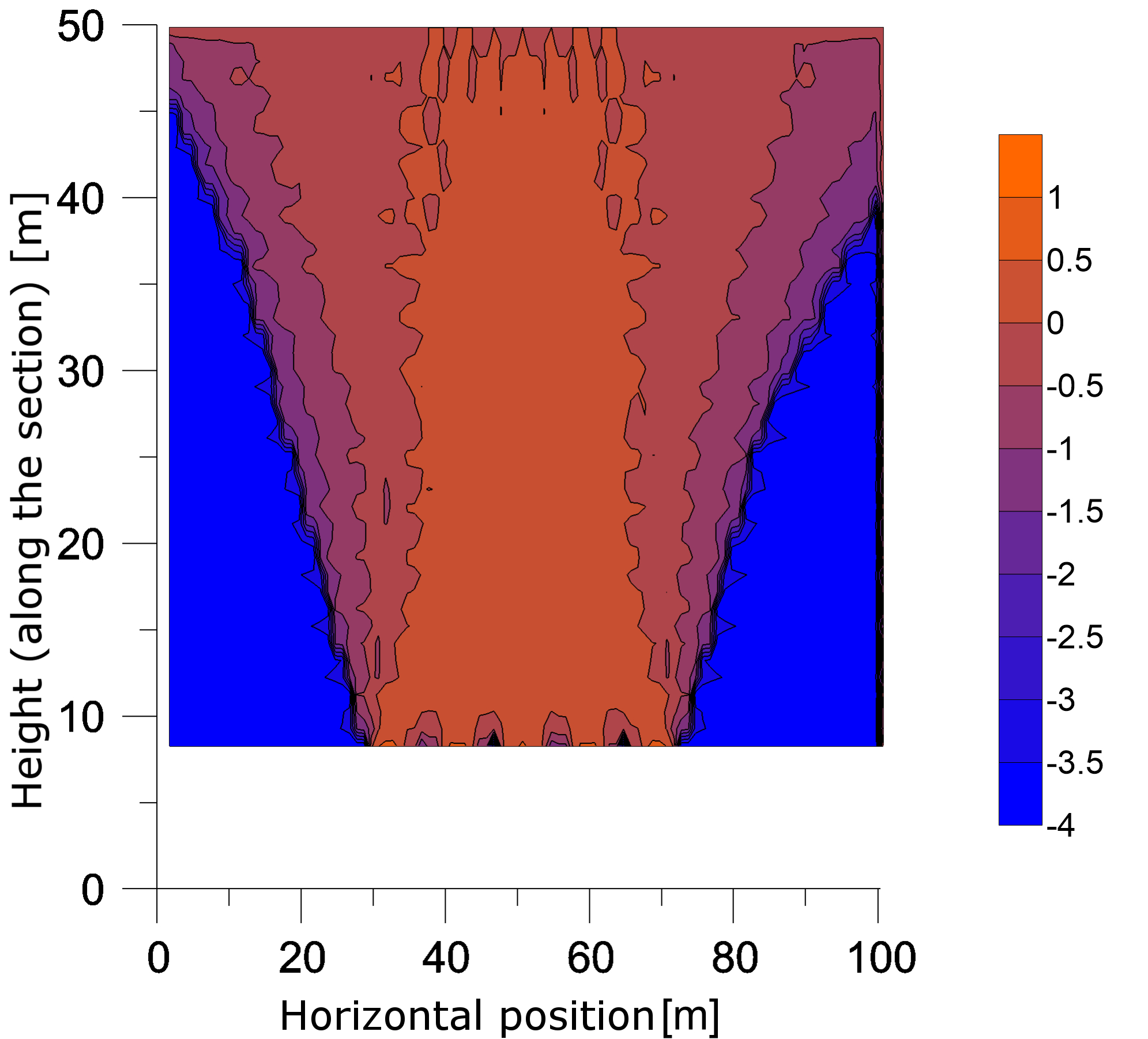} }
    \caption{
The sensitivity map (the logarithms of back projected measurement weight from Eq. \ref{eq:omega}, $\log (w_k)$) of different measurement configurations introduced on Fig. \ref{fig:1}. The scale of axes are not the same.
}
    \label{fig:2}
\end{figure*}

Based on the known top surface (topography) or possible underground structures of the domain under investigation, the efficiency and focal area of a given measurement system can be investigated prior to the measurements, and the parameters of the measurement system can be optimized. In this chapter such a study is presented mainly with the aim to verify performance and understand the details of the output. In this example a predefined 3 m diameter cavity (zero density) is located in a homogeneous measurement environment at 20~m depth and viewed by intuitively organized series of measurements 50~m below a flat surface. Figure \ref{fig:1}. shows the mapping method in two similar measurement configurations (wider and narrower range of positions), while on Figure~\ref{fig:2} the focal range of the mapping can be analyzed based on the projected weights.
%

As can be seen from Figure~\ref{fig:2}, with differently tilted detectors the mapping is slightly differently focused, but the coverage is still weak at the edges. 

\begin{figure*}
    \centering
    \resizebox{1\textwidth}{!}{
    \includegraphics[height=5cm]{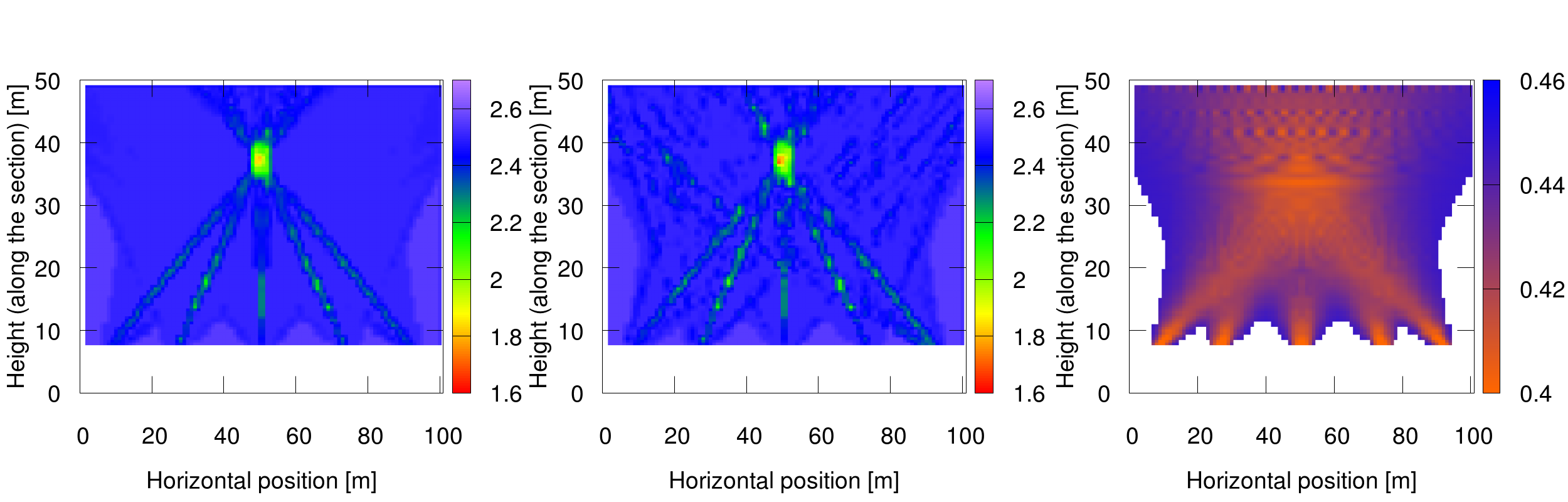}}
    \caption{
Density inversion results with a hypothetical void in a homogeneous rock: without noise (left), with noise if 1 month measurements assumed (middle), and the standard deviation of estimated densities (right) in g/cm\textsuperscript{3}. The scale of axes are not the same.
}
    \label{fig:3}
\end{figure*}

The inversion results are shown in Fig.~\ref{fig:3}, assuming density value $\boldsymbol{\rho}^{(0)}$ = 2500~kg/m\textsuperscript{3} and variance $\sqrt{\boldsymbol{C}_\rho^{(0)}}$ = 450~kg/m\textsuperscript{3}. The left panel is a noise-free (infinite statistics) calculation, which show characteristic artifacts along the projection lines. Mathematically these are due to nonzero off-diagonal elements of $\textbf{R}$ matrix. In terms of the Bayesian approach, one can understand that the {\it a priori} assumption is a flat density map, therefore the most probable {\it a posteriori} density map does not fully describe the zero density anomaly (cavity). The fit fills the cavity, and balances with reduced density along the observation lines. One must note that the artifacts do not reach the magnitude of the real anomaly, but still allow a qualitative verification. Generally the artifacts are linked to the real cavity as a characteristic radial patterns.

A key issue with real measurement is the finite statistics, that is, noise. Inversion with simulated noise (statistical uncertainty after collecting 1 month data, assuming a 0.16~m\textsuperscript{2} detector with 1\textdegree angular bins) are shown in the middle panel of Fig.~\ref{fig:3}. The qualitative picture does not change, and at the cost of the artifacts, the Bayesian approach suppresses oscillations in the inversion result which would result from a standard maximum-likelihood fit.

\section{Inversion of field measurements} 

\subsection{Geometrical configuration}

\begin{figure*}
    \centering
    \resizebox{1\textwidth}{!}{
    \includegraphics[width=6cm]{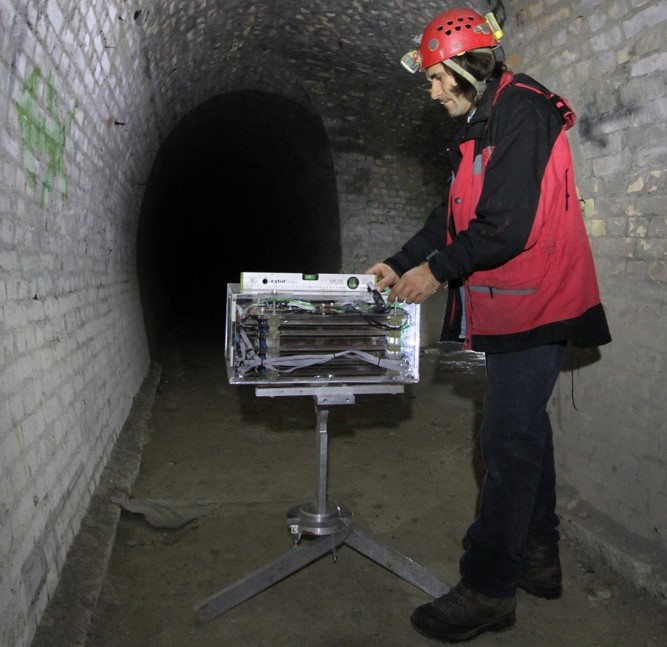}
    \includegraphics[width=6cm]{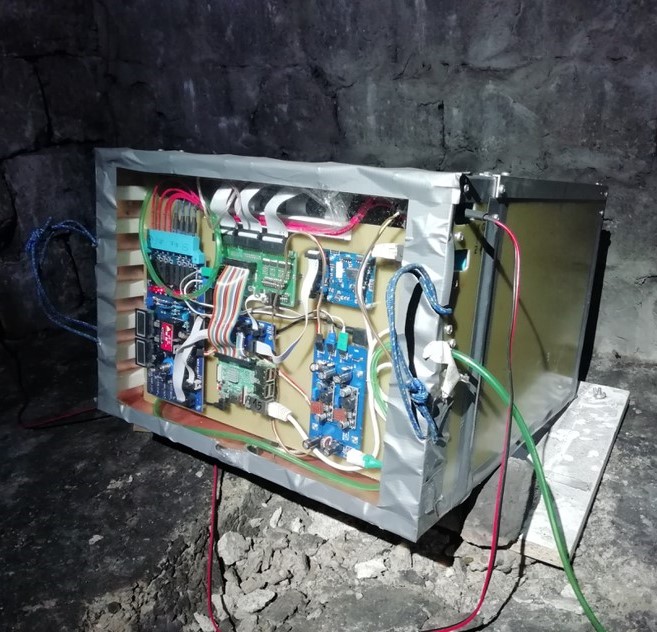}
    }
    \caption{
Images of the tracking detectors consisting Close Cathode Chambers, installed in the Királylaki tunnel.
}
    \label{fig:CCC}
\end{figure*}

The application of Bayes inversion is demonstrated on a real one-line field measurement in the Királylaki tunnel near Budapest, Hungary. Seven measurements have been performed by Close Cathode Chamber technology~\citep{Olah_2012, Gusty_2012, Varga_2013} along the tunnel
(see Fig.~\ref{fig:CCC})
using a 0.16~m\textsuperscript{2} detector with 1\textdegree~angular bins. The detector was placed in the straight tunnel to study the density distribution of the overlying rocks for the purpose of cavity exploration. 

\begin{figure*}
    \centering
    \resizebox{1\textwidth}{!}{
    \includegraphics{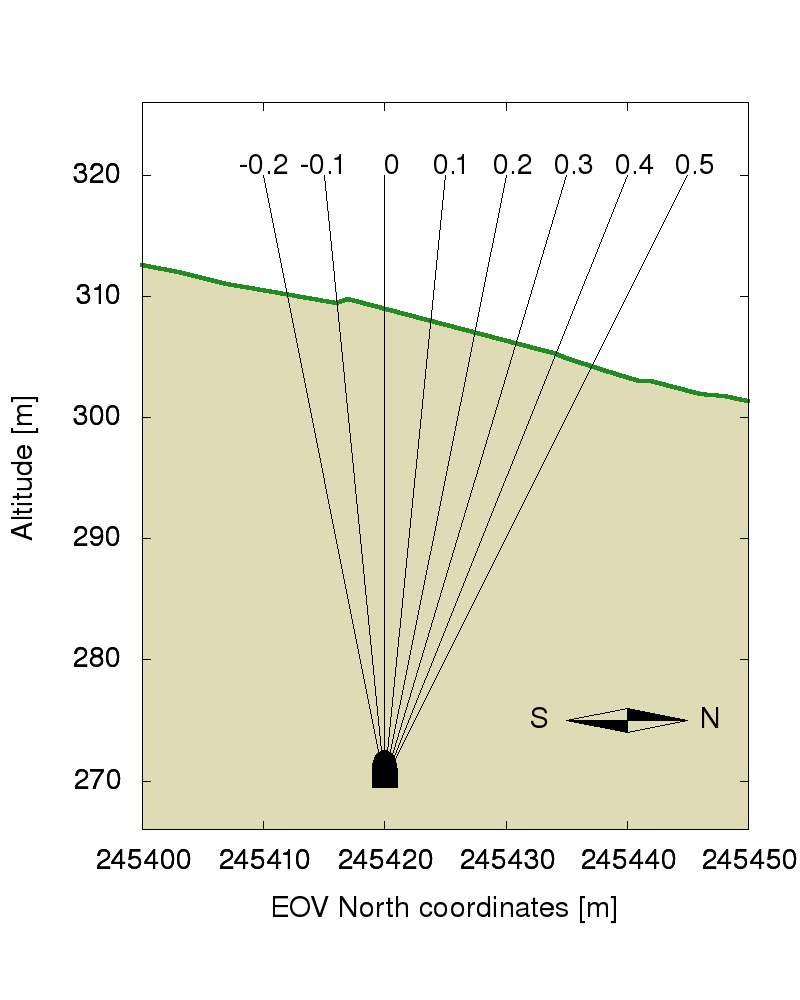}
    \includegraphics{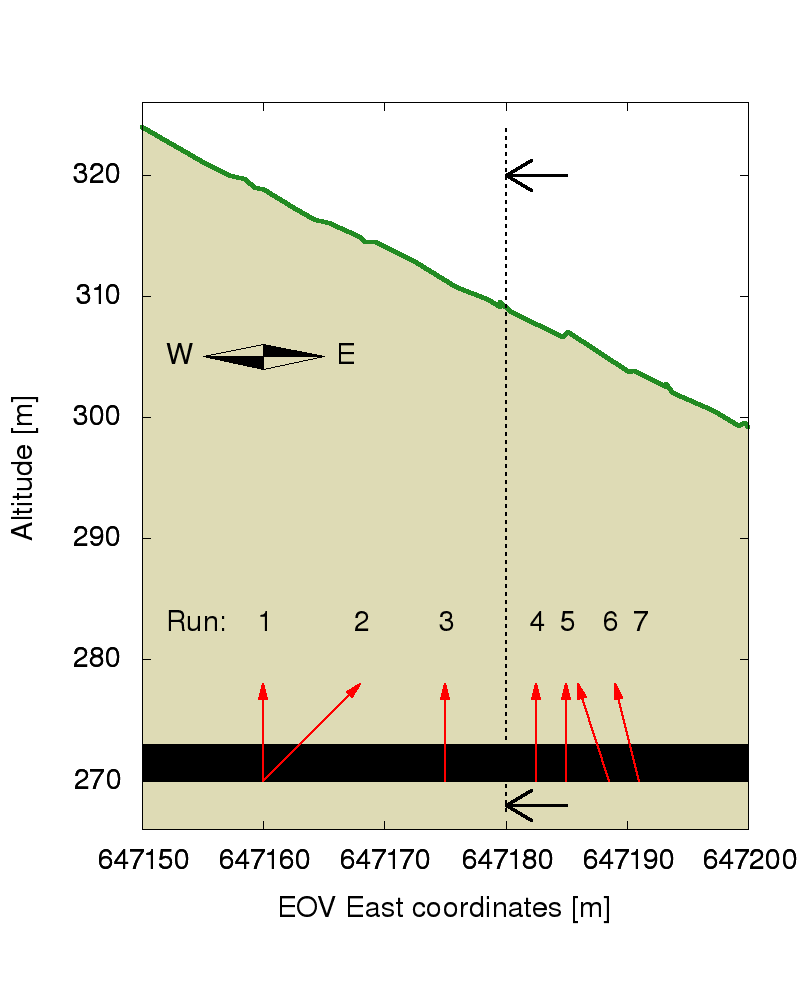} }
    \caption{
Topography and the geometric conditions of the measurements and inversion. Left: the section of the topography perpendicular to the tunnel. The slices (indicated from -0.2 to 0.5 by the tangent of their zenith angles) are the 2D planes in which the tomographic inversion has been made (see Sec. \ref{2+1D}). Right: the section of the topography parallel to the tunnel. Red arrows show the detector pointing of the measurements, dashed line indicates the section of the left figure. 
}
    \label{fig_TOPO}
\end{figure*}

The measurements were taken over a period of approximately one month per position to reach the proper variance and detection limit. Fig.~\ref{fig_TOPO} shows the topography of the examined region and the geometric configurations of the measurements, where the coordinates are on the basis of the National Hungarian Grid (EOV). Run numbers and associated arrows indicate each detector position and viewing direction, collecting data in a $\sim$90\textdegree  cone in each viewing direction. These specific configurations have been chosen based on preliminary surveys to focus on the most significant and closest density anomalies.

\begin{figure*}
    \centering
    \resizebox{1\textwidth}{!}{
    \includegraphics{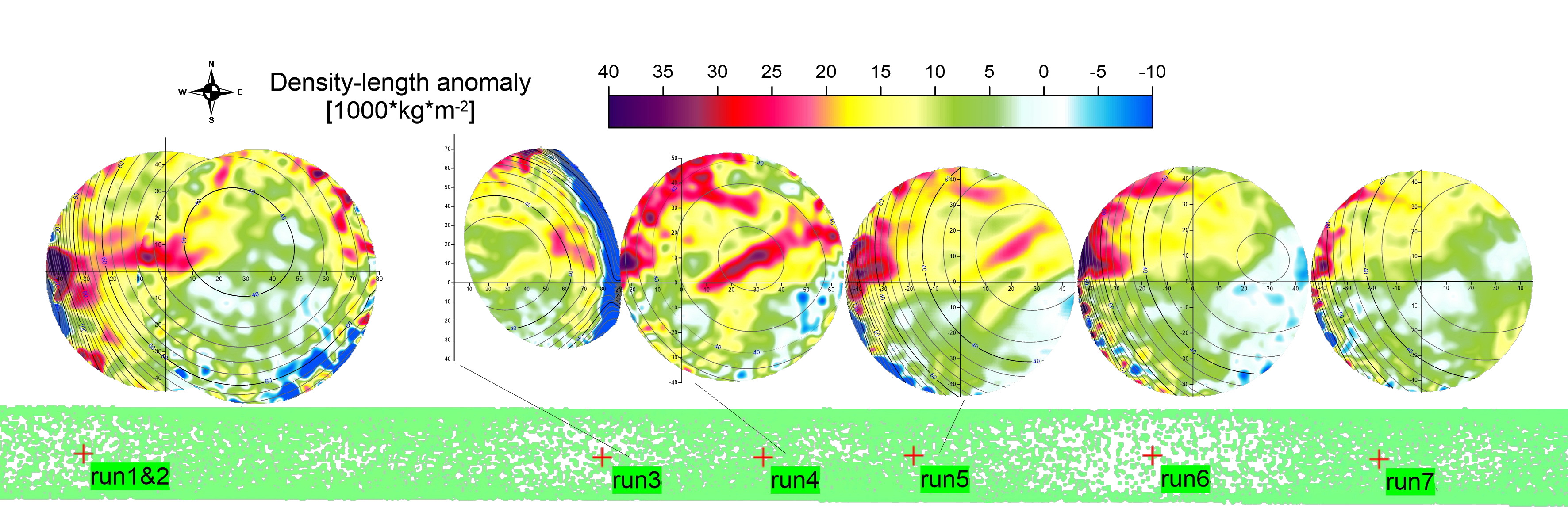} }
    \caption{
The detected density-length anomalies (deviation from a homogeneous density) are shown at the measurement positions in the tunnel, each plotted on a square grid of horizontal coordinate system. Contour lines represent the detector-to-surface rock thicknesses from the given point of view.
}
    \label{fig_meas_series}
\end{figure*}

The detected density-length anomalies (difference between the measured density-length and that of a homogeneous rock) are presented in Fig. \ref{fig_meas_series}. The measured density-lengths were the input for the inversion.

\subsection{Results of inversion and quantification of uncertainties}

\begin{figure*}
    \centering
    \resizebox{1\textwidth}{!}{
    \includegraphics[width=5cm]{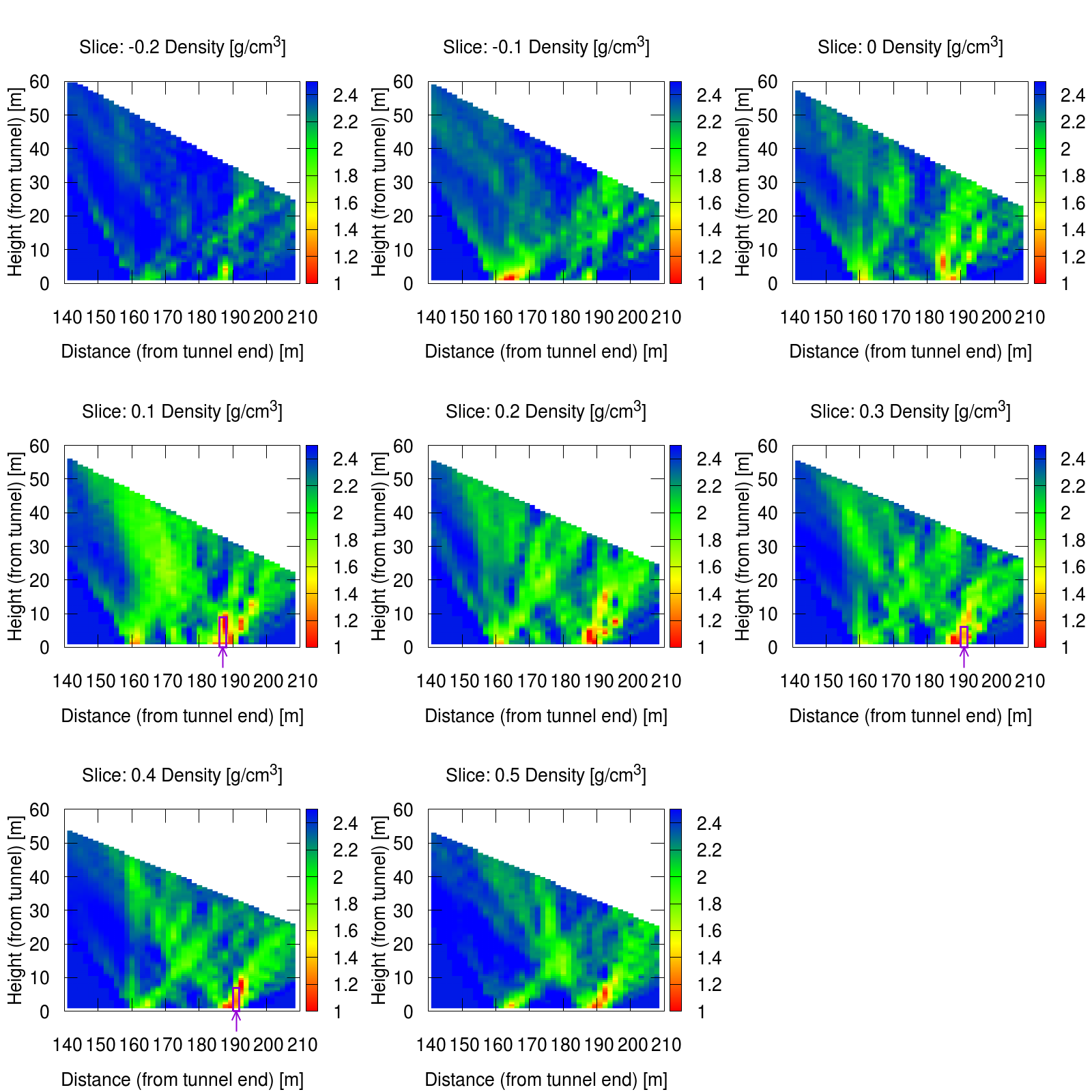} }
    \caption{
The result of the tomographic inversion ({\it a posteriori} density distributions) is shown in relevant slices (as Fig. \ref{fig_TOPO}). On the 0.1, 0.3, and 0.4 slices purple arrows show the validation drill locations.
}
    \label{fig:ALL_inversion_slices}
\end{figure*}

The results of the Bayesian inversion are displayed on the Fig.~\ref{fig:ALL_inversion_slices} in each slice with 1.5 m resolution in horizontal and 0.5 m resolution in vertical direction, assuming density value $\boldsymbol{\rho}^{(0)}$ = 2400~kg/m\textsuperscript{3} and variance $\sqrt{\boldsymbol{C}_\rho^{(0)}}$ = 450~kg/m\textsuperscript{3}. Note that this Bayes-prior variance is larger than the typical measurement error (5\%), therefore gives higher weight to the measured data. The coordinates have been shifted: vertical origin is at the level of the top of the tunnel, the horizontal origin is fixed at the end of the Királylaki tunnel (EOV East 647000). The southern slices (-0.2 and -0.1) do not contain significant anomalies, accordingly, the resulting density distribution is essentially homogeneous in the middle part (in the focus area). The north slices show density anomalies reaching very close to the tunnel. 

\begin{figure*}
    \centering
    \resizebox{1\textwidth}{!}{
    \includegraphics[height=5cm]{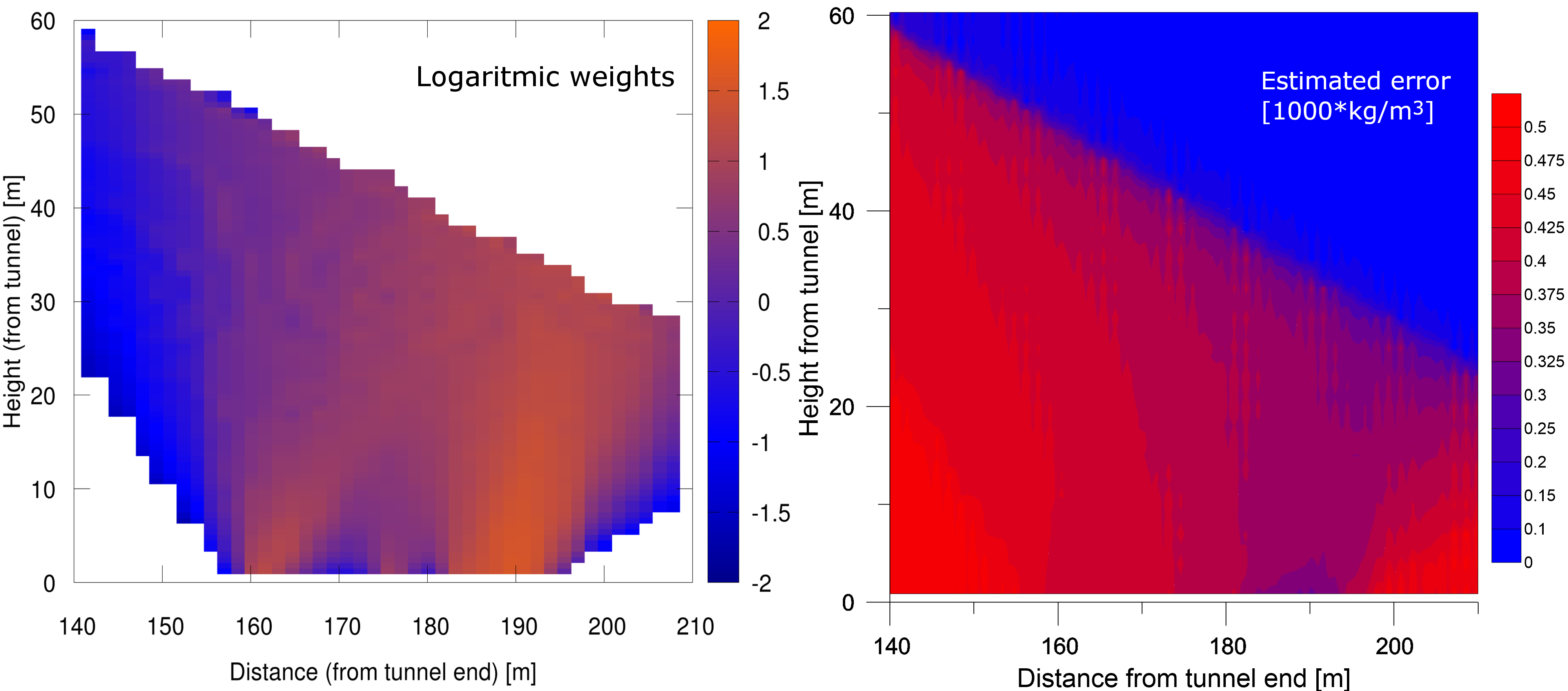}
    }
    \caption{
Maps of quantities describing the focus area and uncertainties in a chosen 2D slice (0.2 tangent slope). Left panel shows the logarithmic weight factors (sensitivity map) which relates to the sum of count numbers from all the measurement lines crossing the given cell. Right panel shows the estimated errors, propagated by the bias calculations.
}
    \label{fig:significance}
\end{figure*}

The inversion works efficiently in the focus area defined by high values of back propagated weight (left panel of Fig.~\ref{fig:significance}). However, this quantity falsely implies high sensitivity close to the detectors as well, since a voxel will be taken into account by multiple measurement line in this nearer region. The standard deviation of the density values (similar in all slices) are mapped in the right panel of Fig.~\ref{fig:significance}. 

\begin{figure*}
    \centering
    \resizebox{1\textwidth}{!}{ 
    \includegraphics[height=8cm]{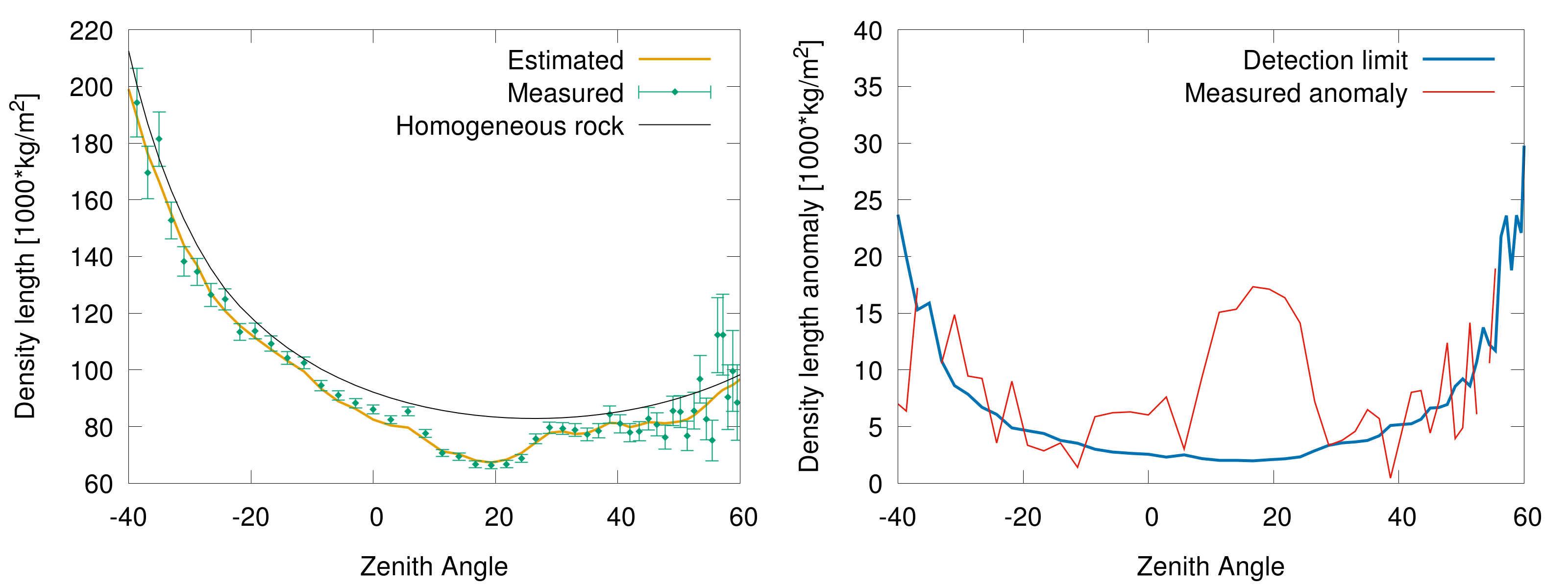}
    }
    \caption{
Left: The density-lengths for a selected measurement (Run 5, -0.1 tangent slope), the result of fitting quality, and the assumed density-lengths for homogeneous rock. Right: The density-length anomalies compared to the detection limit (95\% confidence level).
}
    \label{fig:parameter_fitting}
\end{figure*}

The proper parameter fit in the measurement space is illustrated in Figure~\ref{fig:parameter_fitting}, where the data for a selected detector location are compared with the density-lengths calculated from the fitted densities. The asymmetry of the curves can be explained by the topography. The quality of the fit is similarly excellent at all measurement points. The standard deviation is minimal in the direction of the focus range. With these measurements the problem of detectability is also demonstrated: the anomaly compared to the theoretical measurements calculated with the reference rock density can be detected at the selected 95\% confidence level (1.65 sigma).

\begin{figure*}
    \centering
    \resizebox{1\textwidth}{!}{
    \includegraphics[height=5cm]{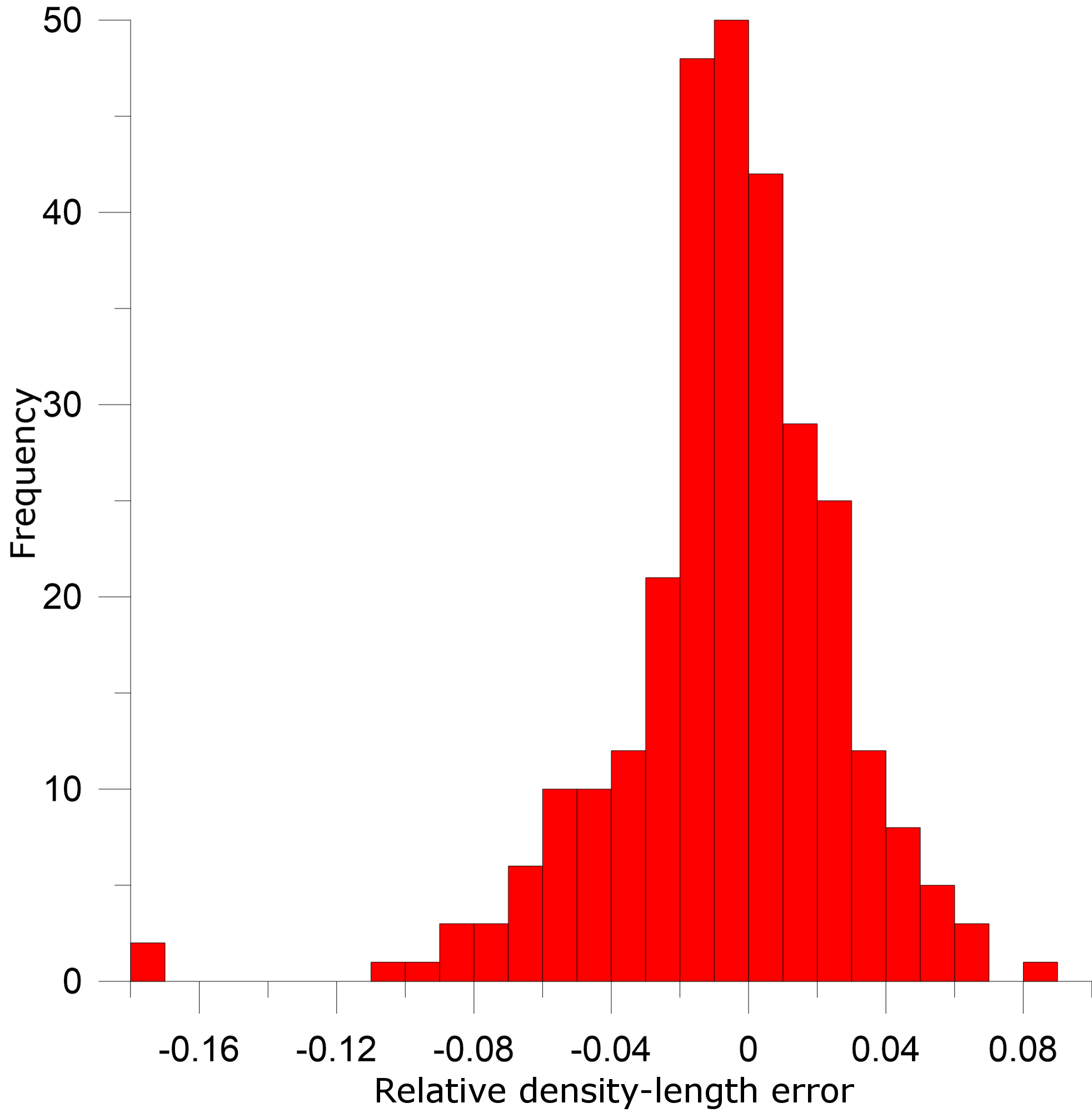}
    \includegraphics[height=5cm]{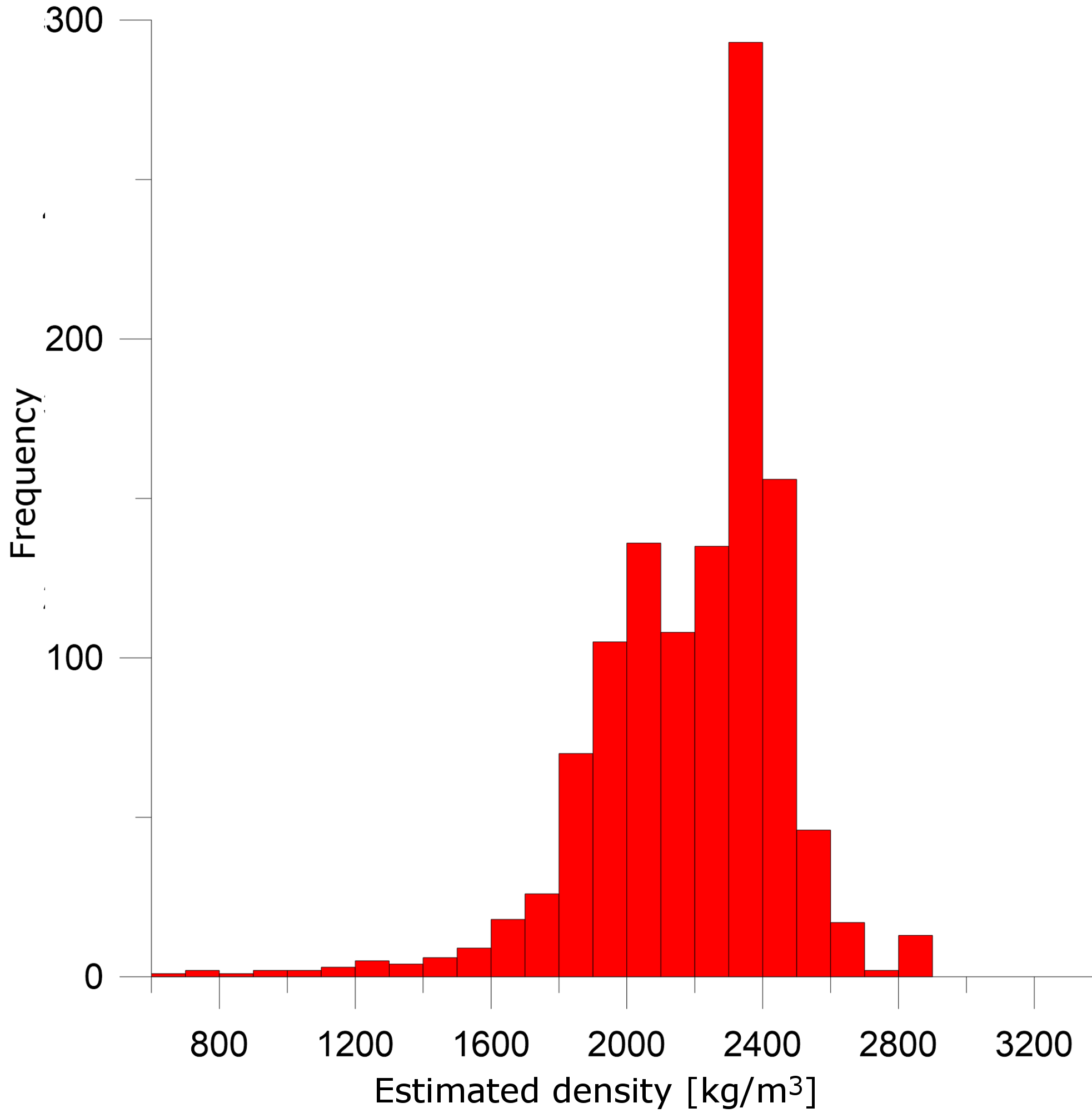} }
    \caption{
Left: The histogram of estimated error (residuals) for fitted density-length data (ratio of estimations and measurements). Right: The histogram of estimated densities (right).
}
    \label{fig_histogram}
\end{figure*}

The histogram (Fig.~\ref{fig_histogram}) shows that the estimated error distribution (residual distribution) has an almost zero mean Gaussian shape distribution. The bias due to the Bayesian term is hardly noticeable. This may be explained by the relatively small fraction of fractured or cavity zones. The fitted parameters (density distribution) show a more complicated distribution, with the peak at the Bayesian prior of the homogeneous assumption.

\section{Validation using drill core samples}


The measurements, both viewed as transmission images or tomographgic results, indicated large anomalies -- decreased density regions -- which are promising targets for a direct verification. Drilling seemed possible from the inside of the tunnel, with positions and directions aiming for anomalies close to the tunnel ceiling.

\begin{figure*}
    \centering
    \resizebox{1\textwidth}{!}{
    \includegraphics[width=12cm]{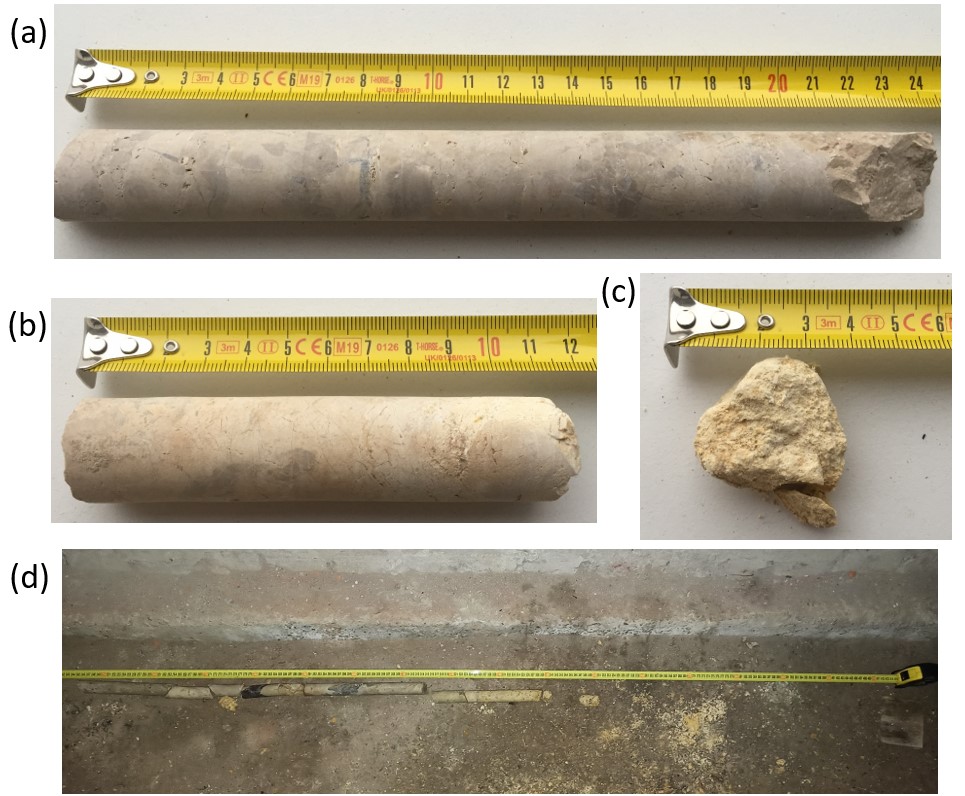} 
    }
    \caption{
Drill cores with different densities: (a) intact cherty dolomite (2.6-2.7 g/cm\textsuperscript{3}); (b) slightly altered dolomite close to the walls of the fissures (2.4-2.5 g/cm\textsuperscript{3}); (c) altered dolomite powder (less than 1.8 g/cm\textsuperscript{3}); (d) the full extent of the last 2 m drill core from one of the drill hole. Dolomite powder only partially recovered.
}
    \label{fig_photos}
\end{figure*}

The length of the control drill holes were limited to 10~m due to technical reasons. Altogether, three drill holes were bore between the zenith angle of 3\textdegree and 22.5\textdegree with the length of 5.4~m, 5.8~m and 9.2~m, indicated in Fig.~\ref{fig:ALL_inversion_slices} with purple arrows in slice 0.1, 0.3, and 0.4. Although none of the drills found empty voids except the 20--50 cm space between the brick wall of the tunnel and the original rock body, the validation of the results were relevant. The low density zones detected by the measurements were large fissures filled up with significantly low density ($\sim$1.8 g/cm\textsuperscript{3}) altered dolomite powder, while the density of the intact rock (cherty dolomite) was 2.6-2.7~g/cm\textsuperscript{3} (see in Fig. \ref{fig_photos}). The contact boundaries of the high and the low-density zones (the walls of the fissure) were at the position previously predicted by the muon measurements by 20~cm accuracy. One of the drill holes went completely through of one of the low-density zones reaching the further wall of the fissure. Most of the dolomite powder has been washed away by the water of the diamond core driller, making it difficult to even continue the drilling operation. The extent of the low-density zone was the same as predicted. Generally, despite the differences between the predicted and the actual absolute densities measured on the drill cores, the overall geometrical structure of the fissure zones were detectable by muon tomography at very high precision.

\section{Conclusions}

Bayesian inversion tuned with homogeneous {\it a priori} density map proved effective in solving the largely one-sided tomographic problem of an actual multi-view muography measurement: the relevant Bayesian assumption made the inverse sufficiently stable. The performed measurements applying a gaseous moun tracking detector system was originally with the aim of search for unknown cavities or density anomalies. High quality data were taken in the artificial Királylaki tunnel, in Budapest, Hungary. In parallel to the muon tomography measurements, complex geophysical cross-check measurements were done: mapping, scanning, and after evaluating the results, drilling for rock samples.

The angular resolution of the trackers, better than 1\textdegree, enabled a spatial resolution of 1-2~meters to meet the needs of cavity exploration. Owing to the stability of the inversion and the resolution achieved, a sufficiently detailed estimate of the 3D distribution of crack zones in the rocks under investigation was obtained in a way that it could be verified by drilling. 


One must note that the parameter bias due to the application of Bayes' principle is largest in the region of the cavities, or anywhere departing from the \textit{a priori} input, because the Bayes-approach "prefers" a solution not too far from the assumption. Anomalies will not disappear due to this effect, but density contrast reduces. Once data suggests the existence of anomalies, then more focused, better positioned, or higher statistics confirmation measurements may be planned, to improve the quantitative evaluation. The principal merit of the Bayesian approach is that it offers a controlled and properly formulated method to extract the information from the limited available information.

\section*{Acknowledgements}

This work has been supported by the Joint Usage Research Project (JURP) of the University of Tokyo, ERI, under project ID 2020-H-05, 
the "INTENSE" H2020 MSCA RISE project under GA No. 822185, 
the "Mine.io" HEU project under GA No. 101091885, 
the Hungarian NKFIH research grant under ID OTKA-FK-135349 and 
TKP2021-NKTA-10, 
the János Bolyai Scholarship of the HAS and 
the ELKH-KT-SA-88/2021 grant. 
Detector construction and testing was completed within the Vesztergombi Laboratory for High Energy Physics (VLAB) at Wigner RCP.

L. Balázs conceptualized the mathematical formalism, calculated the inversion results of the synthetic and field measurement, created figures, and wrote the first version of the draft. 
G. Nyitrai formalized research goals and ideas, planned and carried out the muographic survey and validation drills, produced data, created figures and texts, reviewed the manuscript.
G. Surányi formalized research goals, provided tools and study materials, planned and carried out the muographic survey and validation drills, created figures and texts, and reviewed the manuscript.
G. Hamar formalized research goals, produced data, and acquired funds.
G. G. Barnaföldi formalized research goals, planned and carried out the muographic survey and validation drills, and reviewed the manuscript.
D. Varga formalized research goals and ideas, supervised and reviewed the formalism and the manuscript, added texts, and acquired funds.

\begin{dataavailability}
The data underlying this article will be shared on reasonable request to the corresponding author.
\end{dataavailability}

\end{document}